# Fast-staged CNN Model for Accurate pulmonary diseases and Lung cancer detection

Abdelbaki Souid
*MACS Laboratory, RL16ES22, ENIG, University of Gabes, Gabes, Tunisia,*
SouidAbdelbaki@gmail.com

Mohamed Hamroun
*3IL - XLIM - University of Limoges, France.*
mohamed.hamroun@xlim.fr

Soufiene Ben Othman
*PRINCE Laboratory Research, ISITcom, Hammam Sousse, University of Sousse Tunisia .*
ben_oth_soufiene@yahoo.fr

Hedi Sakli
*EITA Consulting, 5 rue du chants des oiseaux, 78360 Montesson, France.*
saklihedi12@gmail.com

Mohamed Naceur ABDELKARIM
*MACS Laboratory, RL16ES22, ENIG, University of Gabes, Gabes, Tunisia.*
naceur.abdelkrim@enig.rnu.tn

*Abstract*— Pulmonary pathologies are a global health concern, often leading to fatal outcomes when not promptly diagnosed and treated. Chest radiography is a primary diagnostic tool, but the availability of experienced radiologists is limited. Artificial Intelligence (AI) and machine learning, particularly in computer vision, offer a solution. This research evaluates a deep learning model for detecting lung cancer, specifically pulmonary nodules, and eight other lung pathologies using chest radiographs. The study combines diverse datasets, comprising over 135,120 frontal chest radiographs, to train a Convolutional Neural Network (CNN). A two-stage classification system, incorporating ensemble methods and transfer learning, triages images into Normal or Abnormal categories and identifies specific pathologies, including lung nodules among eight conditions. The deep learning model achieves impressive results in nodule classification, with a top-performing accuracy of 77%, sensitivity of 0.713, specificity of 0.776 during external validation, and an AUC score of 0.888. While successful, some misclassifications, primarily false negatives, were observed. In conclusion, this model demonstrates the potential for generalization across diverse patient populations due to the geographic distribution of the training dataset. Future improvements may involve integrating ETL data distribution and adding more nodule-type samples to enhance diagnostic accuracy.

*Keywords*— *Pulmonary nodule; Deep learning, Transfer learning, medical image processing, Lung diseases, Frontal chest x-ray Introduction.*

## I. INTRODUCTION

Chest radiography is widely recommended and routinely utilized as the primary imaging modality to detect suspected respiratory illnesses, particularly lung cancer. The popularity of this investigation can be attributed to its procedural simplicity, widespread availability, easy accessibility, and minimal radiation exposure when compared to chest CT (Computed Tomography) scans. The simplicity of the procedure enables early detection of lesions, thereby facilitating prompt disease assessment and management[1]. Based on numerous studies, the implementation of chest radiography in lung cancer screening has been shown to improve survival rates. For instance, Strauss et al[2] conducted randomized controlled trials that presented compelling evidence indicating that CXR screening can enhance lung cancer survival by detecting tumors at an earlier stage[3]. Moreover, a large population-based cohort study demonstrated that CXR screening decreased lung cancer mortality by 18% in high-risk individuals. Additionally, a case-control study revealed that CXR screening reduced lung cancer mortality by over 20% [4]. hese studies highlight the significant potential of CXRs in lung cancer screening.

However, identifying nodules in chest imaging is considered a challenging image classification task. Lung nodules are low-contrast tissues that are difficult to distinguish, and early studies have utilized pre-defined features extracted from image processing or manually hand-crafted to extract meaningful features such as shapes, textures, densities, and intensities, which are then classified using machine learning techniques [5]. Nevertheless, such feature engineering tasks have limited generalization capabilities. Until recently, computing power and the availability of large data sets allowed this approach to flourish, these advances in deep learning and large data sets allow the algorithm to match the performance of medical professionals in various fields such as biomedical fields [6, 7], medical imaging tasks, including the detection of diabetic retinopathy [8], skin cancer classification [9], and lymph node metastases detection [10]. Consequently, the automatic diagnosis of chest imaging has garnered increasing attention [11], Artificial intelligence and machine learning applications in the medical field have expanded significantly and now include sensitive medical tasks such as skin melanoma[12], brain tumor[13] and other pathology detection models [14–16].

This work presents a two-stage deep learning pipeline that is designed to identify the presence of abnormalities in patient exams, predict potential pulmonary pathologies from a set of eight possibilities, and localize the identified pathology in the exam image for better explainability. The pipeline utilizes advanced machine learning techniques to achieve these goals, offering a powerful tool for clinicians and researchers in the field of pulmonary pathology identification and diagnosis.

The main objective of this study is to investigate the challenge of recognizing eight types of malignant lung conditions such as Mass and Nodule. The model consists of two blocks: the first is a feature extractor, which includes a pre-trained Convolutional Neural Network (CNN) that uses state-of-the-art techniques, while the second block applies transfer learning to refine the model's predictions. The study is divided into several sections, beginning with an introduction, followed by a review of previous work on the classification of medical data in section 2. Section 3 presents the suggested approach, while section 4 discusses the outcome of the experiment that was carried out. This work is concluded in the final part of the study.

## II. RELATED WORK

The ability of machine learning algorithms, especially deep learning, to identify abnormalities in X-ray images has grown in popularity in recent years. Artificial intelligence is used in medical research to aid in diagnosis, and some studies have shown positive and accurate results. Strategies used by previous researchers to manage lung diseases using artificial intelligence and deep neural networks are discussed here.

The majority of patients that are diagnosed with lung cancer are in the mature stage with a lack of a good prognosis. Aside from the late staging of diagnosis, the variability of imaging features and histology in lung cancer makes it difficult for doctors to select the optimal treatment approach [17] The imaging features of lung cancer vary from a single tiny nodule to ground-glass opacity, multiple nodules, pleural effusion, lung collapse, and multiple opacities [18]; simple and small lesions are extremely difficult to detect [19], table 1 illustrates some of these works.

DL-based models have also shown potential for detecting nodules/masses on chest radiographs [20–23], with sensitivities ranging from 0.51-0.84 and a mean number of FP indications per image (mFPI) of 0.02-0.34. Furthermore, radiologist performance at identifying nodules was improved with these CAD models than without them [20].

The topic of Triage to Normal/Abnormal is a commonly studied subject in medical imaging. Studies in this area aim to distinguish normal chest x-rays (CXRs) or prioritize urgent/critical cases to reduce the radiologist's workload or improve the reporting time. To develop a triaging workflow, Tang et al [24] compared the performance of various deep learning models applied to several datasets to distinguish abnormal cases.

A large proportion of the studies use pre-trained standard architectures that can easily be found in deep learning libraries. These architectures are commonly Resnet [25], DenseNet [26], MobileNet [27]. The choice of model depth (such as ResNet-18, ResNet-50, DenseNet-121, DenseNet-161) also varies between studies as there is no standard in this design choice. Most of those studies do not introduce methodological novelty but report or compare the performances of multiple architectures on a given task.). Some studies bring methodological novelty by making use of methods that are known to work well to improve model performance elsewhere. For example, it is known that an ensemble of many models improves performance compared to a single model, some studies that make use of this method are Souid et al [23], Sakli et al [6], those models aim to improve performance and add localization capabilities to an image level prediction model. The work of Bharti et al [28] propose a combination of Conv NN and Spacial Transformer network, the method achieves 73% of accuracy. The research presented by Boyang and Wenyu [29] illustrates the efficiency of multi-scale adaptive residual network to identify four pulmonary pathologies. The model achieves 0,97 AUC.

The work presented in this paper proposes a robust method for detecting anomalies in chest X-rays and identifying multiple lung nodules and masses associated with pulmonary diseases. This is achieved through the use of a convolutional neural network on multiple stages. Additionally, data imbalances are addressed through transfer learning and weighted loss to optimize models for anomaly detection, multiclass classification, and specifically lung cancer prediction. The proposed method offers promising results in the accurate detection of abnormalities in chest X-rays and holds great potential for improving the diagnosis and treatment of pulmonary diseases, particularly lung cancer.

## III. METHODOLGY

The presented work employs a modular two-stage detection system that leverages IoT and edge computing technologies for the analysis of chest X-rays. The initial stage focuses on anomaly detection, while the subsequent stage involves detecting eight distinct lung pathology classes using multiple weights. A crucial aspect that enables the success of this work is the utilization of a large dataset, as deep learning models tend to excel when trained on extensive data. figure 1 provides an overview of the IoT and edge computing architecture used in this application. The architecture encompasses interconnected devices, sensors, and edge nodes that facilitate real-time data processing and analysis at the edge of the network, minimizing latency and optimizing resource utilization. This architecture enhances the efficiency and effectiveness of analyzing chest X-rays by enabling rapid and localized data processing

*A. Overview*

EfficientNet is a family of efficient Convolutional Neural Networks (CNNs) designed to improve accuracy while reducing the number of parameters, computation, and memory usage required to perform image classification tasks. The main concept behind Efficient-Net is to balance the scaling of network dimensions which leads to improved accuracy and low computational complexity. These models have been proven to achieve state-of-the-art performance on various benchmark datasets for image classification tasks, as well as being capable of adapting to other vision-related tasks [30]. The presented work uses EfficientNetB0 [30] for the abnormality triage, also the transfer learning strategy is been used in this work, figure 2 illustrates the architecture including the training procedure which will be further explained in next section.

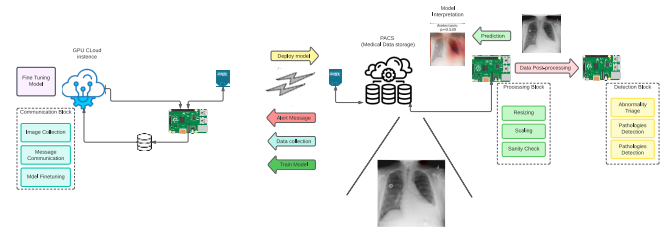

Fig. 1. Presented solution full architecture: Edge communication for data acquisition, model training, and model's inferencing

The model pipeline, as depicted in figure 1, is capable of processing both PACS data streams and individual samples. The first model within the pipeline examines the presence of abnormalities in the provided CXRs. In case abnormalities are identified, the second model is triggered for further analysis. Upon completion of the pipeline cycle presented in the figure 2, the model generates predictions related to abnormality and various pathologies, with a particular focus on lung cancer, including masses, nodules, and seven other pathologies. This approach exhibits promising outcomes in accurately identifying abnormalities and pathologies related to the lungs, thereby showcasing significant potential for enhancing the diagnosis and treatment of pulmonary diseases.

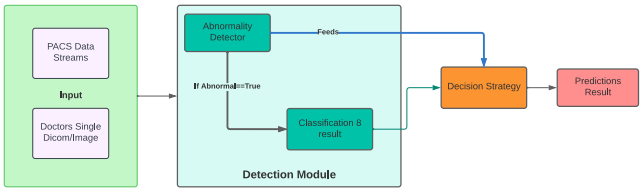

Fig. 2. Proposed model architecture and workflow: End to End pipeline, starting with the input where the data is injected either from a PACS system (in batches) or by samples, then the detection module where the two model receive the image, and we end up with prediction (in the decision strategy).

### B. Dataset presentation

Training abnormality detection models or any deep learning models require a proper architecture and an important number of samples as a set of data. For this task, chosing to work with the VinDr-CXR by Ha Q. Nguyen et al. [31], This dataset is open source and consists of 18,000 chest radiographs in total, subdivided into the training set and test set, and text mined with fourteen disease image labels. This dataset is extremely useful for calibrating the number of distributions of certain classes. For the second model, eight classes of lung diseases are considered in this paper. Wang et al. [32] proposed Chest-x-ray 8 one of the largest open-source data sets by comprise 108,948 frontal view chest radiographs collected from 32,717 unique patients with eight disease image labels. Furthermore, Wang et al. [32] expanded the Chest-x-ray 8 to include 112,120 scans, and 14 pathologies labels that were text mined. They included "No Finding" for samples that did not have an illness. "Infiltration, Mass, Nodule, Pleural Thickening, Atelectasis, Cardiomegaly, Consolidation, Edema, Effusion, Emphysema, Fibrosis, Hernia, Pneumonia, and Pneumothorax" are the diseases. However, this dataset is weighted with several faults, beginning with a large class unbalance caused by an unmatched number of normal samples. A second issue is a large number of uncommon classes, such as Hernia and Pneumonia, which include fewer than a thousand examples. combing two other datasets to address these problems, the first dataset is the OTC and Chest-X-ray dataset by KS. Daniel et al. [33], comprises two sets one for the OCT and the other chest-x-ray, our interest is focused on the chest-x-ray set, which comprises 5,863 chest radiograph text mined in 2 categories "Normal and Pneumonia", as it is clear by now our aim is from using this dataset is to boosting the distribution frequency of Pneumonia class. Then used the VinDr-CXR by Ha Q. Nguyen et al. [31], which is implemented for training the first model. There is still some drawback for this method as it is not possible to leverage the metadata contained in both: Vin-Dr-CXR and Chest-x-ray 14 and OTC and Chest-x-ray dataset due to big differences in the metadata starting the data type: Vin-Dr-CXR and Chest-x-ray 14 and OTC and Chest-x-ray dataset owing to significant changes in metadata beginning with the data type: The Vin-Dr-CXR comes in DICOM format, but the other two datasets come in JPEG format, which is a significant difference that can't be rely on. Second, as previously stated, the Vin-Dr-CXR and chest-x-ray 14 share some of the listed pathologies but not all of them, so choosing between using the full dataset and having 17 classes or reducing the number of pathologies. For this work, acquiring the following eight lung diseases: "Atelectasis, Cardiomegaly, Consolidation, Nodule/Mass, Pleural thickening, Pneumothorax, Pulmonary fibrosis, and Pneumonia.

### C. Dataset processing and augmentation

Each model required a specific data processing methodology, even the datasets have a relatively common pathologies category. Starting with the general abnormality model, originally the dataset comprises 15 classes with a relativity small data distribution compared to the NIH dataset, which is convenient in our case for the model to be familiar with the maximum number of cases. The dataset was provided in the DCIOM format which added some additional time to process the data and explore any utility of the meta-data stored within the DCIOM files. To reduce the classes dimension, a PCA was applied. One other important side was the clarity of the scans which was absent in some samples, addressing this issue by implementing the Contrast limited adaptive histogram equalization (CLAHE) [34], Hence using transfer learning-based architecture to enhance model performance, the training data were converted into channels instead of one channel, also training deep learning requires heavy computer computation and time consumption, hence reducing the size of the images from (512, 512) to (224, 224).

For the second model, applying the required data cleaning such as detecting the existence of outliers, preventing data leakage, and sanity check for each class. Then, the reformulated dataset is split into three subsets the biggest one for training, the second one the validation, and the last small subset used to test the model performance.

Data augmentation is a valuable technique for reducing model overfitting by increasing the amount of data training and introducing various distortions and noise to the training data. On photos, using four types of data augmentation techniques: horizontal flip, brightness, and contrast, random-sized crops, and, normalizing the pixel values of images so the range of pixel intensity values is between 0 and 255.To address the data unbalanced, we had modified the across-entropy loss present in equation 1 by adding a positive wight and a negative eight.

$$L^w_{cross-entropy} - (w_p \log(1 - f(x))) \qquad (1)$$

Where the positive weight $w_p$ multiplied by the positive frequency of each class $freq_p$ to be equal to the negative weight $w_n$, multiplied by the negative frequency $freq_n$ of each class presented in the formula

$$w_p \times freq_p = w_n \times freq_n \qquad (2)$$

And N is the total number of patients.

### D. Pipeline models architectures

Both models chares some similar specification, the two proposed model uses the EfficientNet, however each model have his own role in the all-in pipeline.

- **EfficientNet Architecture**

The EfficientNet [30] is a collection of models based on the baseline network described in table 1. Its main component is the Mobile Inverted Bottleneck Conv (MB-Conv) Block, introduced in [27] and shown in Fig 1. The EfficientNet Convolutional Network family is based on the idea of starting with a high-quality yet compact baseline model and uniformly scaling each of its dimensions systematically with a fixed set of scaling coefficients, both of the presented model uses EfficientNet CNN as a base. the time complexity of the Convolutional layer is $O(k.n.d^2)$, However, as mentioned prior, EfficientNet use NAS architecture (Neural Architecture

Search) and also added classification layers, so time complexity becomes not easy to calculate.

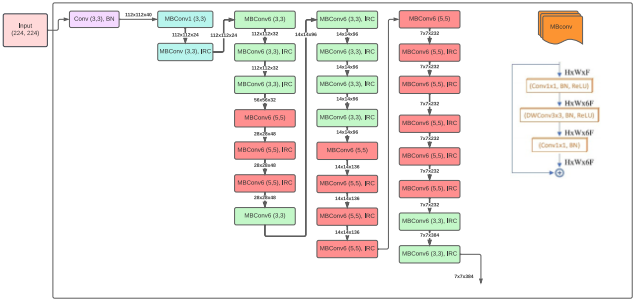

Fig. 3. EfficientNet architecture: MBConv stands for Depthwise Conv, 1x1/ 3x3 defines the kernel size, BN is the batch norm, H, W, F means tensor shape (height, width, depth), and ×1/2/3/4 is the multiplier for the number of repeated layers.

The EfficientNet is a very robust Convolutional Neural Network that focuses on scaling which a significant because scaling help improve model efficiency. EfficientNet uses Neural Architecture Search (NAS) that optimizes for both accuracy and FLOPS cost. For the presented work, the Efficient Net (B1, B0) had been used.

The presented text describes the architecture of the abnormality detection model using the EfficientNet B0 as a feature extractor. The model architecture consists of a flattened layer, two dense layers with a rate of (256, 128), separated by dropout layers, and a classifier layer with two outputs. Figure 3 is provided to illustrate the model architecture.

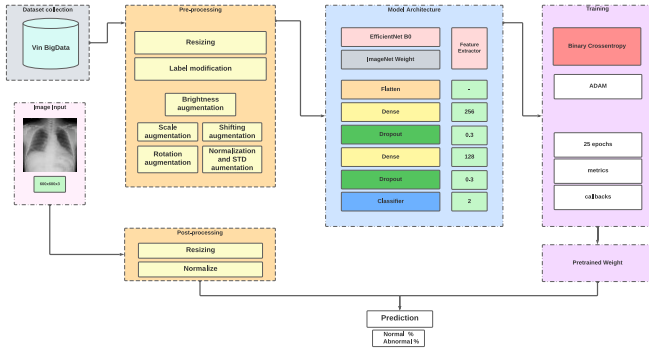

Fig. 4. general abnormality detector architecture: Input 224x224x3, EfficientNet B0 weight pre-trained on ImageNet followed by Flatten layer, Dropout with 0,3 and 2 Dense layers with 128 and 256 and a classifier layer (Dense layer with 2 class).

In order to create a classifier for the eight lung diseases, the EfficientNet B1 was chosen as the feature extractor due to its effectiveness in image classification tasks. Rather than training a new model from scratch, the pre-trained weights of a previous-ly trained network were used to accelerate the learning process. The proposed meth-od consists of two blocks, as shown in Figure 4. A transfer block was added, which included a zero-padding layer to smooth the results, a Convolutional layer with a kernel of 512 and stride of 33, a Global Average Pooling layer (GAP) with a Condi-tional Dropout layer, a Fully Connected layer with 1024 output nodes, and a classifi-er with 8 output nodes. The dataset was split into 70% for training, 20% for valida-tion, and 10% for stage 1. The input size for the EfficientNet B1 was set to (224,224, 3). The Adam optimization algorithm was used to train the model, starting with a learning rate of 0.1e-2, which was gradually reduced to 0.01e-3 using the ReduceLROnPlateau scheduler.

The model was initially trained by fixing all the layers of the pre-trained model and training it for 15 epochs, where each epoch had half the number of steps. This was followed by a second training phase, where the model was fine-tuned for anoth-er 15 epochs. An examination was conducted on the 0.5 threshold to improve accu-racy and enable the model to generalize better.

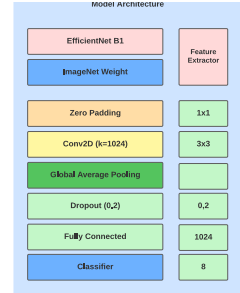

Fig. 5. the eight classes classifier model architecture: Input 224x224x3, EfficientNet B1 weight pre-trained on ImageNet followed by 1 zero-padding layer, 1 Conv layer with k3,3 also a GAP layer, Dropout with 0,2 and 1 Dense layer with 1024 and a classifier layer (Dense layer with 8 class).

### E. Modules Evaluation

Both experiments were evaluated using receiver operating characteristic. To measure the quality of predictions of the presented models, it is highly relevant to calculate the evaluation metrics namely the sensitivity, specificity, accuracy, and AUC of ROC (Area Under the Curve Receiver operating characteristic). All the training and testing are made on Google Colab [35] instance with 1 V-GPU with 15 GB of RAM and 1 CPU for 16 GB of RAM:

$$precision = \frac{TP}{TP + FP} \quad (3)$$

$$Sensitivity = \frac{TP}{TP + FN} \quad (4)$$

The TP, FP, and FN refers to True Positive, False Positive, and False Negative. The F1-score is calculated using the following formula:

$$F1 - score = 2 \times \frac{precision \times recall}{precision + recall} \quad (5)$$

## IV. EXPERIMENTAL RESULT AND DISCUSSION

### A. Results Evaluation

Reviewing the model's performance revealed some notable results. The first model was trained on 20 epochs with Adam optimizer and categorical cross entropy loss, viewing the model accuracy curve presented in the figure 5 we notice some sign of plausible overfitting, in general the model accuracy got 94% during training/validation and 93% on 3000 external test set.

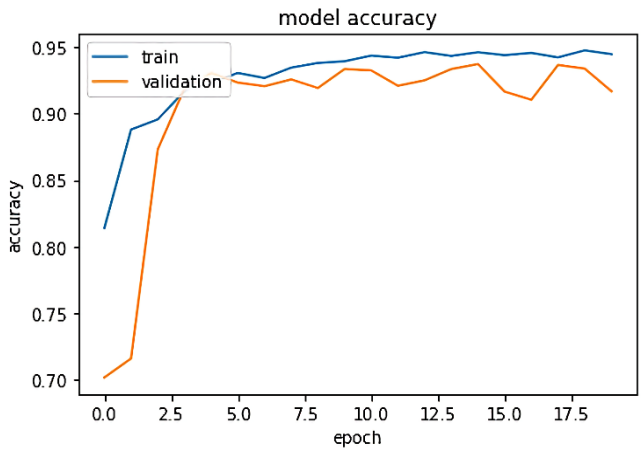

Fig. 6. Accuracy Value in training/validation phase: blue curve train, orange curve validation.

To further investigate the model performance, we calculate the confusion matric illustrated in the figure 6. The results of the classification were very acceptable, 246 images over 3000 images were miss classified. Also, calculating the precision, recall and F1-score based on the test data, the model achieves 93% of precision for normal and abnormal cases, the model also achieved the highest F1-score for the abnormal cases with 95%, table 2 illustrate the described results. Other metrics results were presented in the table.

TABLE I. THE ABNORMALITY TRIAGE MODEL RESULTS

| Metrics | Normal | Abnormal |
|---|---|---|
| Accuracy | 0.937333 | 0.937333 |
| F1-score | 0.889542 | 0.956259 |
| Negative predictive value | 0.935792 | 0.941542 |
| Positive predictive value | 0.941542 | 0.935792 |
| Precision | 0.941542 | 0.935792 |
| Sensitivity | 0.842984 | 0.977640 |
| Specificity | 0.977640 | 0.842984 |

In our assessment, we discovered a large dispersion of findings compared to AUC values. In addition to the loss function, this might result from the random initialization of the models and the stochastic nature of the optimizer.

The EfficientNet B1 indicates that having more features improves total model performance based on the loss value. The performance metric on EfficientNet B1 during training shows 0.265 in training and 0.245 invalidations for Loss, Accuracy with between 76% and 95% class-wise, 81,5% for sensitivity, 80,8% for Specificity, and 0,888 for AUC, table 1 shows the individual results, and Fig 7 shows the Receiver Operating Characteristic for the AUC.

TABLE II. EVALUATION METRICS REPORT FOR THE EIGHT PATHOLOGIES PREDICTION MODEL

| Pathologies | P | R | F1-score | mAP@0.5 |
|---|---|---|---|---|
| Atelectasis | 0.764 | 0.831 | 0.759 | 0.872 |
| Cardiomegaly | 0.858 | 0.859 | 0.858 | 0.926 |
| Consolidation | 0.764 | 0.82 | 0.762 | 0.861 |
| Nodule/Mass | 0.771 | 0.713 | 0.776 | 0.835 |
| Pleural thickening | 0.744 | 0.786 | 0.742 | 0.841 |
| Pneumothorax | 0.824 | 0.879 | 0.822 | 0.925 |
| Pulmonary fibrosis | 0.792 | 0.809 | 0.791 | 0.882 |
| Pneumonia | 0.951 | 0.825 | 0.954 | 0.963 |

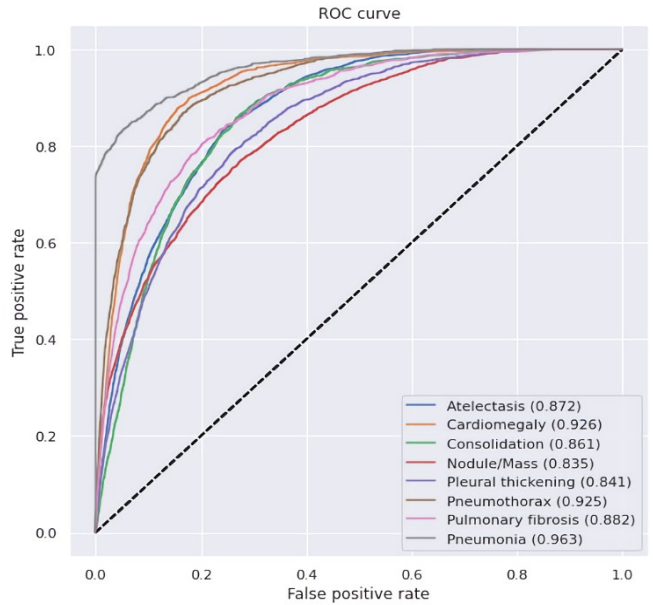

Fig. 7. Receiver Operating Characteristic Area Under Curve for the EfficientNet B1: The Legend bottom right distinct each class with a color. The highest AUC value for pneumonia was 0.963, lowest Mass/Nodule is 0.835.

*B. Discussion*

Overall, Overall, the combination is very promising for real-world accurate prediction, meaning that if the second model did not detect the correct abnormality, the pipeline is at least capable to recognize whether abnormality exists or not which is the role of the first model. For the second model, according to the obtained results from the evaluation process, we would like to say that the model gets good performance toward the detection specific classes that are presenting Lung cancer: the mass and Nodule, the had proven its efficiency to detect lung nodule/mass w To begin, we compared the obtained results to those by Wang et al [32]. While DenseNet 121 have a superior individual AUC in 8 out of 8 disease classes, EfficientNet B1 has a 30% improvement in more than three of them. Baltruschat et al [36] are among those who have contributed to this study. While our model was trained with fewer epochs, it nevertheless produced good results for "Atelectasis, Cardiomegaly, Pleural thickening, Pulmonary fibrosis, and Pneumonia" as well, a slightly higher average AUC of 0.888. The EfficientNet B1 is more generic; three classes surpass the 0.9 AUC score, and the GradCam [37] Visualization is quite precise even for picture resolutions of 224,224, the figure 9 illustrate some of the test examples. The EfficientNet B1 also achieves extremely acceptable results when compared to other state-of-the-art studies, such as Wang et al [32] with an AUC value of 0.745 and Baltruschat et al [36] with an AUC of 0.806, as well as table 3 present individual AUC diseases comparison.

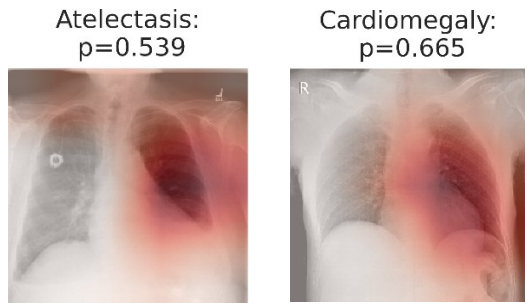

Fig. 8. Gradient Class activation map applied on some of the test samples: image on the right: GT: Atelectasis, prediction; Atelectasis, second sample GT: Cardiomegaly prediction: cardiomegaly, prediction correct even if the probability is low.

When applying the Gradient class activation map, the model does a decent job by highlighting the region of interest correlated with the prediction label, the model is capable of localizing the predicted pathology even with low prediction probability, this indicates how far the presented model can explain the prediction also the fairness of the model.

| Pathologies | Wang et al, [23] | Baltruschat et al [36] | Proposed |
|---|---|---|---|
| Atelectasis | 0,7 | 0,755 | **0,872** |
| Cardiomegaly | 0,81 | 0,875 | **0,926** |
| Consolidation | 0,703 | 0,749 | **0,861** |
| Nodule/Mass | 0,693 | 0,821 | **0,835** |
| Pleural thickening | 0,669 | 0,761 | **0,841** |
| Pneumothorax | 0,799 | 0,846 | **0,925** |
| Pulmonary fibrosis | 0,786 | 0,818 | **0,882** |
| Pneumonia | 0,799 | 0,714 | **0,963** |
| Average | 0,744875 | 0,792375 | **0,888125** |

CONCLUSION

Based on the findings, we believe that our deep learning-based CAD might help radiologists improve the accuracy of their diagnosis of a wide range of pulmonary lesions. In an emergency situation where, expert radiologists are either overworked or unavailable, a deep learning system, for example, can speed up image interpretation. Although it can improve interpretation accuracy, our method should only be used to supplement diagnoses of pulmonary problems.

We found also that implementing 2 stage detection the radiologist has less chance to skip pulmonary abnormalities, also the model not only can detect the listed pathologies but also can discover rare classes due to these combinations.

Concerning evaluation, the proposed approach brought improvements compared to others' work, with an overall AUC of 0.888 and AUC of lung Mass/Nodule of 0.835, Specificity of 77. 6%, and sensitivity of 71.3%, and with less than 16 epochs.

Our research has significant limitations. To begin with, it did not use the complete dataset, which had over fifteen diseases. Second, no expert comparisons were included in our analysis because it was a feasibility study. Third, the discrepancy between the AUC and the accuracy needs more study, which necessitates the usage of other measures. There were many approaches to handle the imbalanced dataset, and it was necessary to investigate the efficiency differences between the strategies.